\documentclass[a4paper,11pt]{article}
\usepackage{pos}
\usepackage{gensymb}

\title{The X-Ray Binary Population with \textit{Fermi}-LAT}
 \ShortTitle{The X-Ray Binary Population with \textit{Fermi}-LAT}

\author*[a]{Max Harvey}
\author[a]{Cameron B. Rulten}
\author[a]{Paula M. Chadwick}

\affiliation[a]{Durham University,\\
  South Road, Durham, United Kingdom}

\emailAdd{max.harvey@durham.ac.uk}
\emailAdd{p.m.chadwick@durham.ac.uk}
\emailAdd{cameron.b.rulten@durham.ac.uk}

\abstract{Binary star systems represent a significant proportion of the Galactic stellar population, with X-ray binaries being an important subset of these for high energy astrophysics. Although hundreds of X-ray binaries are detected in the Milky Way and beyond, only 12 of these systems are listed in the 4FGL-DR2, the latest \textit{Fermi}-LAT point source catalogue. With such a small number detectable by \textit{Fermi}-LAT, much is still unknown about the mechanisms by which these systems emit $\gamma$-rays. We present the method and current status of our large-scale survey of the X-ray binary population using over 12 years of \textit{Fermi}-LAT data, and current catalogues and background models. }

\FullConference{37$^{\rm{th}}$ International Cosmic Ray Conference (ICRC 2021)\\
		July 12th -- 23rd, 2021\\
		Online -- Berlin, Germany}

\begin{document}
\maketitle

\section{Introduction}
X-ray binaries (XRBs) are star systems in which a compact object (typically a neutron star or a black hole) accretes matter from a donor star \citep{reig_bex-ray_2011} \citep{verbunt_origin_1993}. XRBs are typically subdivided into two classes: the high mass X-ray binaries (HMXBs), which have more massive donor stars and accretion occurs through wind-driven interactions, \citep{walter_high-mass_2015} and the low mass X-ray binaries (LMXBs), which have less massive donor stars and where accretion occurs through overflow of the donor star's Roche lobe \citep{tout_wind_1991}.

The \textit{Fermi} Large Area Telescope (LAT) \citep{atwood_large_2009} has detected $\gamma$-ray emission from a number of binary systems, both XRBs and otherwise, with 8 HMXBs and 4 LMXBs being listed in the latest point source catalogue, the 4FGL-DR2 \citep{abdollahi_fermi_2020} \citep{ballet_fermi_2020}. In addition to these 12, $\gamma$-ray emission from the HMXB microquasar SS\,433 is also observed with \textit{Fermi}-LAT, and well-documented in the literature, although it is not a listed source in the 4FGL \citep{rasul_gamma-rays_2019} \citep{li_gamma-ray_2020}. 

XRBs which emit $\gamma$-rays can be divided into two general classes. The microquasars (Cygnus X-3, Cygnus X-1 and SS\,433 for example) are thought to emit $\gamma$-rays via the acceleration of particles in their jets \citep{dubus_gamma-ray_2015}. The second class, the $\gamma$-ray binaries, require a pulsar accretor as opposed to a black hole and represent most of the $\gamma$-ray emitting XRB population. For an isolated pulsar, the wind from a pulsar spinning-down forms a pulsar wind nebula; however, in a binary system this environment is much denser due to the stellar wind from the companion star. In this case, the particle winds from the two stars result in the formation of a shock front, producing $\gamma$-ray emission which may be modulated with the orbital period \citep{chang_gev_2018}. 

Given that we see $\gamma$-ray emission from only a small handful of the total XRB population, much is unknown about these systems and their emission mechanisms. In this ongoing study, we are undertaking a complete survey of the X-ray binary population of the Galaxy with \textit{Fermi}-LAT, using the catalogues of Liu et al \citep{liu_catalogue_2006} \citep{liu_catalogue_2007} as a basis for our search (with some minor changes to make these consistent with our understanding of the XRB population in 2021). The motivation for this project is to discover any traces of faint, or transient, $\gamma$-ray emission from previously undetected X-ray binary systems in order to increase the number of known $\gamma$-ray emitting XRBs. Given that \textit{Fermi}-LAT provides all sky coverage, and has done for over a decade, the \textit{Fermi} data represents the ideal tool for such a large scale survey.

\section{\textit{Fermi}-LAT Observations}

\subsection{Data Reduction and Analysis Parameters}

Most XRBs are located on the Galactic plane, which is an extremely luminous background source of $\gamma$-rays, overlaid with a densely-packed field of point sources and extended $\gamma$-ray sources. This presents a challenge when attempting to model faint or transient emission nearby as the chances of a false positive result increase greatly. 

We follow a binned maximum likelihood method \citep{mattox_likelihood_1996} in order to model the \textit{Fermi}-LAT data in a region of interest (ROI) centered around each of our XRBs independently (there is considerable overlap between ROIs, but we generate a separate ROI for each XRB, rather than looking for several XRBs per ROI). We use the \texttt{Fermitools} in conjunction with \texttt{Fermipy} \citep{wood_fermipy:_2017} in order to follow a standard data reduction and analysis chain consisting of photon selection, computing instrument exposure and livetime, and finally binning the photons into spatial bins of angular width of $0.1 \degree$, and employing 8 energy bins per decade. Table \ref{tbl:params} gives the analysis parameters used; we used the entire available observation duration from the beginning of the \textit{Fermi} mission to the point where we began our analyses in February 2021 (MET 600307205). 

\begin{table}
\centering
\begin{tabular}{cc}
\hline \hline
Observation Period (Dates) & 04/08/2008 - 05/02/2021 \\
Observation Period (MET) & 239557417 - 600307205 \\
Observation Period (MJD) & 54682 - 58423 \\
Energy Range (GeV) & 0.1 - 500 \\
Data ROI width & $10\degree$ \\
Model ROI Width & $15\degree$ \\
Zenith Angle & $< 90\degree$ \\
Good Time Interval Filter & \texttt{DATA\_QUAL>0 \&\& LAT\_CONFIG==1} \\
Instrument Response  & \texttt{P8R3\_SOURCE\_V2} \\
Isotropic Background Model & \texttt{iso\_P8R3\_SOURCE\_V2\_v1} \\
Galactic Background Model & \texttt{gll\_iem\_v07} \\
Point Source Catalogue & 4FGL-DR2 \\
Extended Source Templates & 8 Year Templates \\
\hline
\end{tabular}
\caption{The parameters used in the likelihood analysis of the region of interest around each X-ray binary system.}
\label{tbl:params}
\end{table}

\subsection{Model Fitting}

We initially fit a model to our data using the \texttt{Fermipy optimize} routine, which iteratively pushes the model parameters close to their maxima. We then free the normalisation of all point sources within $1 \degree$ of the central XRB position, as well as the Galactic and isotropic diffuse backgrounds, and execute a full likelihood fit with the \texttt{MINUIT} optimiser to further improve the fit around the position of the XRBs. To test the accuracy of the model with respect to the data, we generate residual maps of each ROI to ensure that the predicted $\gamma$-ray counts per bin reflect the actual counts in the LAT data.

In order to test for persistent $\gamma$-ray sources in our models we employ log-likelihood hypothesis testing which generates a test statistic (TS) from the ratio of the likelihoods of two hypotheses. This is given in Equation \ref{eqn:TS}. 

\begin{equation}
    \label{eqn:TS}
    \mathrm{TS} = 2 \ln \frac{L(\Theta_{1})}{L(\Theta_{2})}
\end{equation}

From Wilks' Theorem \citep{wilks_large-sample_1938}, for $k$ given degrees of statistical freedom between the two models, the TS equates to a value of $\chi^{2}$, which in turn equates to a $z$-score. For a statistically significant detection of a source, we use the threshold $z = 5 \sigma$, and for a `sub-threshold' $\gamma$-ray excess we take the range $3 \sigma \leq z < 5 \sigma$. We do not include any sources or excesses in our models with $z < 3 \sigma$.

In order to populate our models of each ROI with additional, uncatalogued, sources we run the \texttt{find sources} algorithm in \texttt{Fermipy}. This generates a TS map of the ROI and fits point sources to the four highest TS peaks (where $z > 3 \sigma$)  provided these are more than $0.5 \degree$ away from a more significant peak. After four sources are fitted, this procedure is repeated iteratively a further four times so that up to 20 sources may be added to the model. 

Following this, we check to see if \texttt{find sources} has fitted a source spatially coincident\footnote{Our criterion for spatial coincidence between an added source and the position of an XRB is that the coordinates of the XRB lie within the $95\%$ positional uncertainty radius of the source.}, and if not we add a free source manually to the position of the XRB and refit. We then run a light-curve on the source at the position of the XRB using a 6-month binning scheme. 

In order to identify whether there is some level of $\gamma$-ray emission, whether that is a detection or a sub-threshold excess, we first check if the source fitted to the position of the XRB has a $z \geq 3 \sigma$. This is regarded as evidence for some level of $\gamma$-ray emission from the position of the X-ray binary (but not necessarily from the XRB itself). 

There is strong evidence for variability for $\gamma$-ray variability from XRBs, and hence any light-curve where the flux bins (excluding upper limits where $z < 2 \sigma$) have overall $z$-scores (obtained by multiplying the bin $p$-values together) where $z \geq 5 \sigma$ are also considered to show some evidence $\gamma$-ray emission from the XRB position.  

\section{Preliminary Results}

In order to survey the XRB population in a manageable way, we have first analysed the HMXB population separately to the LMXB population. Approximately 100 sources are included in the \cite{liu_catalogue_2006} HMXB catalogue, including 4 sources in the 4FGL, Cyg~X-3, Cyg~X-1, LS~5039 and LS\,I +61 303. All of these sources are detected more significantly than in the 4FGL-DR2. These sources and their significances are given in Table \ref{tbl:known_binaries}. This increase in significance over the 4FGL-DR2 values is encouraging, as generally doubling observation time (for a steady source) leads to an increase in detection $z$-score of a factor of $\sqrt{2}$, and for 3 of the 4 known binaries (the exception being Cyg X-1) we far exceed the expected increase in $z$-score. Given we increase our increase our observation time by 25\% with respect to the 4FGL-DR2, this suggests an analysis which is improved over that of the 4FGL-DR2, and is likely to detect new $\gamma$-ray sources, whether associated with the XRB population or not.

\begin{table}
\centering
\begin{tabular}{cccc}
\hline \hline
Binary name & TS & $z$-Score & 4FGL $z$-Score \\
\hline
LS 5039 & 18000 & $130 \sigma$ & $62 \sigma$ \\
Cyg X-1 & 88 & $9.4 \sigma$ & $8.6 \sigma$ \\
Cyg X-3 & 860 & $29 \sigma$ & $11 \sigma$ \\
LS I +61 303 & 170000 & $410 \sigma$ & $250 \sigma$\\

\hline
\end{tabular}
\caption{The four 4FGL sources included in the HMXB catalogue with their TS value and corresponding $z$-score from our analysis. For comparison, we include the detection $z$-score for each source provided in the 4FGL-DR2.}
\label{tbl:known_binaries}
\end{table}

We observe some ($z > 3 \sigma$) persistent $\gamma$-ray excess from 16 of the HMXBs surveyed. Of these 16 HMXBs, 5 of these have spatially coincident $\gamma$-ray excesses with significances greater than $z > 5 \sigma$. Additionally, a further 5 ROIs indicate some significant level of transient $\gamma$-ray emission from the position of the binary without persistent excess at the $z > 3 \sigma$ level. 

Simply identifying $\gamma$-ray emission from the position of an XRB is insufficient evidence to claim that an XRB is emitting $\gamma$-rays.  We therefore perform a more detailed analysis of these 21 $\gamma$-ray excesses in order to identify the likely candidate for the emission, XRB or otherwise. The easiest way to identify whether $\gamma$-rays come from an X-ray binary is through correlated variability with activity at other wavelengths, and to this end we generate light-curves using X-ray data from \textit{Swift}-BAT, where such data are available. $\gamma$-ray emission is expected to peak at the orbital periastron of a $\gamma$-ray binary, at the same time as the binary undergoes a Type I X-ray outburst. By binning $\gamma$-ray data by phase and comparing it to the X-ray data, we are able to observe any possible orbital modulation in these systems, which would be a clear indicator that a spatially coincident $\gamma$-ray excess is caused by the XRB itself.  

For this approach to work, we require an accurate orbital ephemeris for the XRB (which generally requires the XRB to have a long orbital period of the order of at least 100 days), or a long term X-ray light-curve. These are not available for many of the systems we include in our survey. In cases where we are unable to examine the orbital phase and X-ray properties of the systems, we perform an `elimination analysis', where rather than attempt to associate the $\gamma$-ray excess with the binary directly, we attempt to prove that it is not likely to be from any other source. We initially localise the position of the $\gamma$-ray emission, thus finding an improved fit to position and reducing the positional uncertainty of the source. If, after localisation, the position of the XRB no longer lies within the positional uncertainty of the excess, we can be confident that this excess is not produced by the XRB. 

As the XRBs are generally located on the Galactic plane, some are located very close to known, bright $\gamma$-ray sources. We  generate light-curves of these nearby objects, and compare enhancements and flares between the nearby sources and the apparent excess from the XRB. If these enhancements are temporally coincident, but brighter, in the neighbouring source, it is likely that the excess is caused by source confusion.

If we localise the $\gamma$-ray excess and it remains coincident with the position of the XRB, and nor do we find any correlated variability between the excess and nearby sources, and the excess lies outside of any extended $\gamma$-ray sources we conclude that the excess \textit{could} represent emission from the XRB, but that there is not enough evidence to claim an unambiguous detection.

\subsection{A False Positive Case Study: GS\,1839-04}

GS\,1839-04 is a HMXB with an unknown accretor and unknown companion star. We observe a $\gamma$-ray excess identified by the \texttt{Fermipy find sources} algorithm spatially coincident with the position of GS\,1839-04 which we designate PS J1842.0-0418. PS J1842.0-0418 has a TS value of 17.8 and an angular offset from the position of GS\,1839-04 of $0.147 \degree$, although the 95\% positional uncertainty around PS J1842.0-0418 is unusually large at $1.01 \degree$. Contained within this uncertainty region are 7 4FGL sources, with the nearest neighbours to PS J1842.0-0418 being the unknown source 4FGL J1842.5-0359c (TS = 318 and an angular offset from the position of GS\,1839-04 of $0.498 \degree$) and 4FGL J1840.8-0453e, the young supernova remnant Kes\,73 \citep{gotthelf_kes_1997} (TS = 1050 and an angular offset from the position of GS\,1839-04 of $0.501 \degree$). Given that the positional uncertainty of PS J1842.0-0418 is so large, encompasses numerous, luminous $\gamma$-ray sources and an extended source, we localise the position of PS J1842.0-0418.

\begin{figure}
    \centering
    \includegraphics[width=380pt]{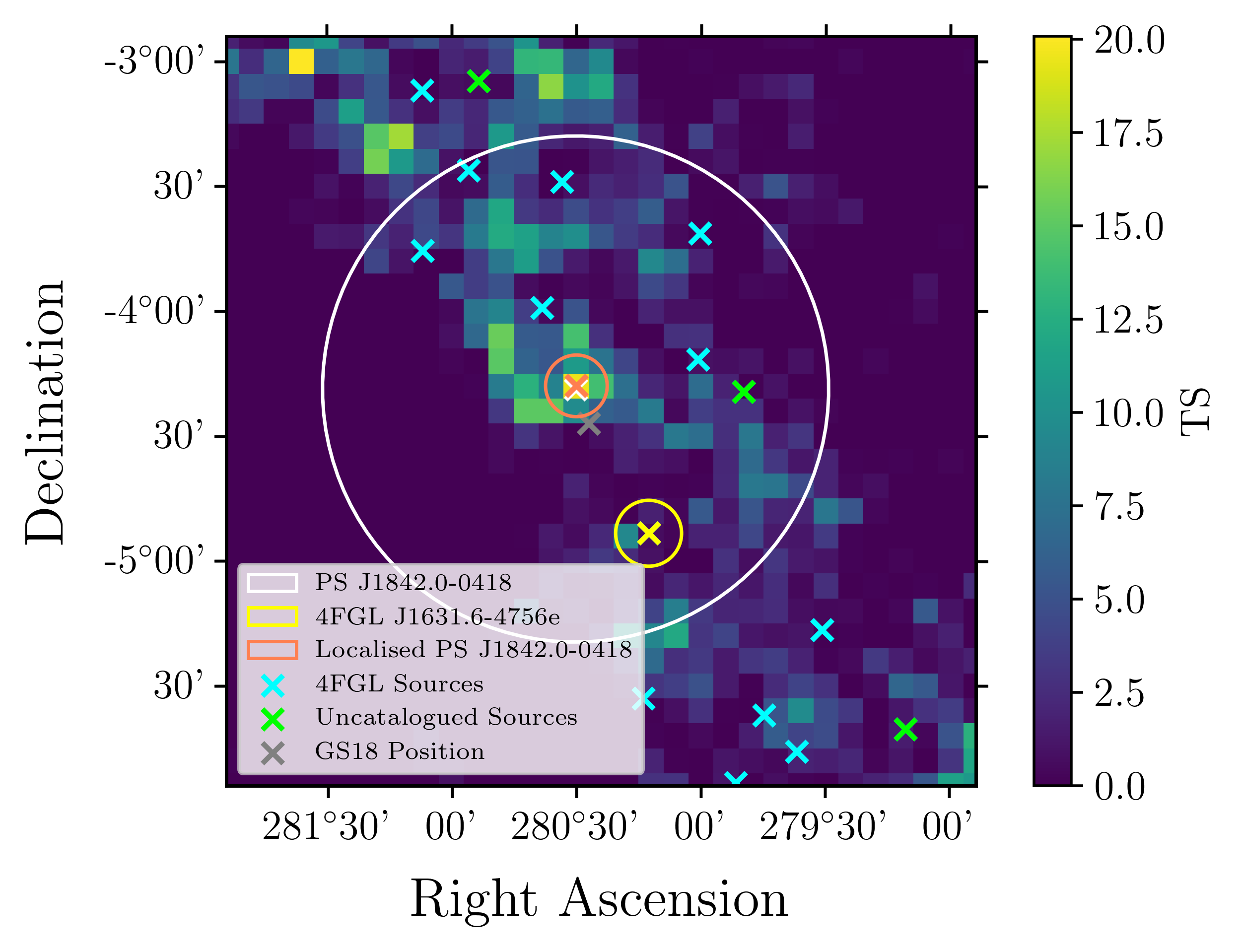}
    \caption{The TS map of the central $3\degree$ of the GS\,1839-04 ROI across the full 12.5 year observation time. Here, the positions of the closest 4FGL sources are indicated in blue, whilst the positions of sources identified with the \texttt{find sources} algorithm are indicated in green. The centroid and extent of the nearby extended source 4FGL J1631.6-4756e is indicated in yellow. The centroid and 95\% positional uncertainty of PS J1842.0-0418 are given in white (before source localisation) and orange (after source localisation). The position of PS J1842.0-0418 barely shifts with localisation hence the orange marker largely overlaps the white one. The position of GS\,1839-04 itself is indicated by the grey cross, and is no longer spatially coincident with PS J1842.0-0418 following localisation.  This TS map is generated after our ROI optimization and fit, but before a point source for GS\,1839-04 is fitted to the model to highlight the spatial coincidence between the excess and the position of GS\,1839-04. Bin widths are $0.1 \degree$ across. }
    \label{fig:GS18_TS}
\end{figure}

Figure \ref{fig:GS18_TS} shows a TS map centered on GS\,1839-04, highlighting the extent of the PS J1842.0-0418 uncertainty and the sources within it. After localising the $\gamma$-ray emission from PS J1842.0-0418 we find that the 95\% positional uncertainty shrinks considerably by approximately an order of magnitude to $0.1234 \degree$. As shown by Figure \ref{fig:GS18_TS}, this means that the X-ray position of GS\,1839-04 is no longer within the 95\% positional uncertainty of PS J1842.0-0418 and therefore PS J1842.0-0418 is very unlikely to represent $\gamma$-ray emission from GS\,1839-04.

\section{Future Work}

As of the time of writing, we are in the final stages of analysing the $\sim$~100-object HMXB population, and our next step is to publish a catalogue of our findings. Following this, our goal is to fully analyse the LMXB population, once again using the \citep{liu_catalogue_2007} as the basis. The LMXB population is likely to be more difficult, as although there are twice as many LMXBs in the Milky Way, there are only half as many in the 4FGL-DR2. These 4 sources are 1SXPS J042749.2-670434, 2S 0921-630, 1RXS J154439.4-112820 and PSR J1023+0038, with detection significances ranging from $8 \sigma$ to $60 \sigma$. Preliminary analysis seems to indicate that a higher proportion of the LMXB population has coincident $\gamma$-ray emission than the HMXB population, however a number of these XRBs are located within globular clusters, many of which luminous $\gamma$-ray emitters. Nonetheless, the prospects of identifying $\gamma$-ray emission from a previously unknown LMXB remains promising.

Beyond the Milky Way, the prospects of detecting $\gamma$-ray emission from XRBs in other galaxies are now real since the discovery of emission from LMC P3,in the Large Magellanic Cloud \citep{van_soelen_orbital_2019}. Following the LMXB survey a logical next step would be to search for extra-galactic $\gamma$-ray emission, likely from the LMC and SMC, both of which are included in \cite{liu_catalogue_2006} and \cite{liu_catalogue_2007}.

\bibliography{references.bib}
\bibliographystyle{JHEP.bst}

%% Full authors list (ONLY FOR COLLABORATIONS)
%\clearpage
%\section*{Full Authors List: \Coll\ Collaboration}
%
%\noindent \textbf{Note comment afterwards:} Collaborations have the possibility to provide an authors list in xml format which will be used while generating the DOI entries making the full authors list searchable in databases like Inspire HEP. For instructions please go to icrc2021.desy.de/proceedings or contact us under icrc2021proc@desy.de.\\
%
%\scriptsize
%\noindent
%first.author$^1$, 
%second.author$^2$, 
%third.author$^3$ % .... more names
%and 
%last.author$^{n}$ \\
%
%\noindent
%$^1$first.affiliation.
%$^2$second.affiliation. % .... more affiliation
%$^{m}$last.affiliation.

\end{document}